\documentclass{sig-alternate}
\usepackage{amssymb}  \usepackage{graphicx} \usepackage{url}

\usepackage[colorlinks=true,linkcolor=blue,citecolor=blue,urlcolor=blue]{hyperref}
\usepackage{algorithm}
\usepackage{algorithmic}
\usepackage{amssymb}
\usepackage{graphicx}
\usepackage{subfigure}
\usepackage{color}
\usepackage[numbers]{natbib}

\newcommand{\mund}[1]
        {{\smallskip\noindent{\bf #1}}}

\makeatletter
\let\@copyrightspace\relax
\makeatother

\begin{document}
%\mainmatter  % start of an individual contribution
\markboth{Class for Lecture Notes in Computer
Science}{Class for Lecture Notes in Computer Science}

\title{Real-Time Bid Optimization for Group-Buying Ads}

\numberofauthors{2}
\author{
\alignauthor
Raju Balakrishnan\\
\affaddr{Computer Science and Engineering} \\
\affaddr{Arizona State University, Tempe AZ USA 85281 }
\email{rajub@asu.edu}
\alignauthor
Rushi P Bhatt \\
\affaddr{Yahoo! Labs}\\
\affaddr{Bangalore, India 560071}
\email{rushi@yahoo-inc.com}
}

\maketitle
\begin{abstract}
  Group-buying ads seeking a minimum number of  customers before the deal expiry are increasingly used by the daily-deal providers. Unlike the traditional web ads, the advertiser's profits  for group-buying ads depends on the  time to expiry and additional customers needed to satisfy the minimum group size. Since both these quantities are time-dependent, optimal bid amounts to maximize profits change with every impression. Consequently, traditional static  bidding strategies are far from optimal. Instead, bid values need to be optimized in real-time to maximize expected bidder profits. This online optimization of deal profits is made possible by the advent of ad exchanges offering real-time (spot) bidding.  To this end, we propose a real-time bidding strategy for group-buying deals based on the online optimization of bid values. We derive the expected bidder profit of deals as a function of  the bid amounts, and dynamically vary bids to maximize profits. Further, to satisfy time constraints of the online bidding, we present methods of minimizing computation timings. Subsequently, we derive the real time ad selection, admissibility, and real time bidding of the traditional ads as the special cases of the proposed method.   We evaluate the proposed bidding, selection and admission strategies on a multi-million click stream of 935 ads. The proposed real-time bidding, selection and admissibility  show significant profit increases over the existing strategies. Further the experiments illustrate the robustness of the bidding and acceptable computation timings.
\end{abstract}

\section{Introduction}

Web based deals offering deep discounts to a group of online buyers on products and services is a fast growing market. Group-buying deals attract new customers as well as guarantee customer traffic within a stipulated expiry date for local businesses like restaurants and tour operators~\cite{dealsWSJ}. Most of these group-buying deals are sold by intermediaries  like Gropon,  Groupbuy and many other daily deal providers.  Though these intermediaries depended on email based marketing models in the past, banner ads in social networking and other sites are increasingly used to attract deal customers.

Unlike the traditional ads,  group-buying intermediaries receive their payment only upon satisfying the minimum number of conversions before the deal expiry (i.e. if the deal tips).  This implies that if the deal does not tip, advertiser  loses the amount used to buy impressions, and receives no payment. If the advertiser fulfills or exceeds the guarantee, he receives a payment equal to the product of number of conversions and pay per conversions---similar to the traditional ads. This model is used by popular group-buying advertisers like Groupon, and Groupbuy among many other deal providers. Though most of these deals tips for sites like Groupon in current email based marketing, tipping the deals will get harder with increasing competition to attract business owners and shift to the display-ad based marketing. The proposed strategy enables the deal advertisers to offer more aggressive tipping points, hence more volume of sales to merchants. Further, this  model is easy scale to other forms of group-buying campaigns---like penalties for not meeting tipping similar to guaranteed display ads.

 To maximize the profits while bidding for group-buy ads, bidders have to minimize cost by bidding low, but still have to win sufficient number of conversions to satisfy guarantees before the deal expiry. Bidding high increases the probability of winning impressions thereby improves the chance of the deal tipping. On the contrary, higher bids increase the payment to the exchange thereby reducing the profit. Hence bids need to be optimized considering these two conflicting pulls.  This maximal profit bidding necessitates dynamic bid optimization based on the time to expiry and the number of received conversions.  We  address this problem of maximizing deal bidder profits, by real-time optimization of bids to minimize the cost of impressions while satisfying the deal tipping guarantees. %Note that though we consider the guaranteed conversions, the problem formulation and the proposed optimal bidding strategy is equally applicable to guaranteed clicks or displays on substituting appropriate probabilities instead of conversion rates.

 For group-buying deals, the traditional static bidding strategies based on optimization of expected profits of a single impression is far from optimal. A significant difference from the traditional ads is that the optimal bid value depends on the time to expiry and number of more conversions required to satisfy the guarantees.  For example, consider a deal requiring just a few more conversions to fulfill the guarantee. If the deal is about to expire the advertiser would have to bid higher amounts to increase the probability of winning more  impressions. On the other hand,  if the time to expiry is long  for the same deal, he would better off bidding smaller amounts winning fewer fraction of impressions to minimize the payment to the exchange (since there would be higher number of user visits in larger time intervals). Evidently, the optimum bid amount is a function of the time dependent parameters  like the time to expiry and the additional number of conversions needed to satisfy guarantees. Due to this time dependence of optimal bids, any static bidding strategy will be sub-optimal, necessitating real-time bidding. Fortunately, this dynamic bid optimization is made possible by the  advent of ad exchanges offering real-time  auctions (e.g. RealMedia, DoubleClick, AdECN).

Since the revenue is conditional upon tipping the deal, the bidding strategies are significantly harder than the traditional non-guaranteed bidding. In addition to the dynamic quantities mentioned above, deal profit depends on a number of static quantities: pay per event, number and bid distributions of other bidders,  conversion rates, and the auction mechanism~\cite{krishna2009auction,easley2010networks}. Consequently, formulating and maximizing expected profits---which is  a function of all these static and dynamic quantities---is significantly harder. Adding to the complexity, the optimization is online necessitating low computation timings.

Our method of optimizing profit for guaranteed deals has two steps: (i) Formulating the expected profit (ii) Maximizing the  profit against the bid.  For the first step, we derive the expected profit as a function of the bid value, time to expiry, fulfilled conversions, amount spent to buy impressions, auction mechanism, click through rate  and the number and distribution of the other bidders. Since many of these parameters are dynamic as described above, the objective function value changes as the bidding progresses. Among all these parameters, the only  parameter the bidder can change is his own bid amount. Hence we optimize the expected profit against the bid amount in the second step. When the profits are optimum, the deal bids are in a symmetric Bayesian Nash equilibrium similar to  the traditional ads~\cite{krishna2009auction,easley2010networks}.

Considering the complexity of the optimization, a closed form solution is unlikely. Though the optimization is against a single variable (i.e. bid amount), our analysis shows that  the objective function is  neither convex nor quasi-convex (unimodal). Consequently,  an optimization method guaranteed to converge to optimal bids on every instance is unlikely. Further, the derivative of the objective function is harder to solve than the objective itself. Considering these factors, we resort to direct numerical optimization (without using gradients) starting from multiple points.

\mund{Running Time Minimization: }Since the optimization is online, computation time needs to be minimized. Therefore we explore running time optimization in multiple levels. Firstly,  we use a fast converging Brent's optimizer.  Secondly,  we reformulate the objective for faster computation for typical parameter values. Further, we approximate  large binomial cumulative probability expressions with a single term normal approximation.  Since the changes in the optimal bids for subsequent impressions are incremental, we reuse optimal bid values of previous impressions whenever changes are likely to be negligible. %Further  we limit random-restarts of the search only to the beginning few cycles.

\mund{Extensions: }Interestingly, the solutions of many related problems can be directly derived from the proposed objective function. We describe the four proposed extensions below:
\begin{description}
\item[(i) Deal Selection]Deal selection chooses the best deals to bid to maximize the bidder profits. Combining optimal bidding and selection, we derive the bidders' private value and the marginal profit increase for the impression for each deal. The deal with the highest marginal profit increase is the greedy optimal selection.
\item[(ii) Deal Admissibility]Admissibility is the problem of predicting whether bidding for a group-buying deal is likely to be profitable based on its attributes.
The intermediary or the advertiser  may decide to accept or reject a deal campaign based on admissibility criterion.  We show that a special case of our objective function combined with the bid optimization provides effective admission control.
\item[(iii) Non-Bidding Selection] For non-bidding scenarios like the publisher directly selecting the deals to display, the proposed formulation suggests optimal selection among the inventory of deals.
\item[(iv) Non-Guaranteed Ads]We show that the real time bid optimization of traditional non-guaranteed ads is a special case of the proposed optimization. When  there are no guarantees, the proposed objective function reduces to expected profits of traditional ads, yielding known optimal static bid formulations. Thus the method serves as a unified real time bidding strategy for both guaranteed and non-guaranteed ads. %For non-guaranteed ads in static environments, both the proposed real time bidding and the current static bidding are equally optimal. However, unlike the static bidders,  real time bidder can adapt  to changing parameters like click through rates and competitor bid distributions in dynamic environments~\cite{chen2011real}.
\end{description}

\mund{Evaluations and Results: }We evaluate the proposed methods and the extensions in a query log of size 9.3 million impressions of 935 ads. In our first set of experiments, we compare our  profits of the proposed real time  strategy with the optimal static and base adaptive baselines. The results show that the proposed strategy improves the profits over the baselines significantly.  Subsequently, experiments showing improved profits in spite of violated assumptions of the competitor bids demonstrate robustness of the strategy. Further, our running time evaluations demonstrate acceptable optimization timings. Finally, our evaluations of the ad selection and admissibility demonstrate that the extensions improve profits significantly over the baselines.

Rest of the paper is organized as follows. The next section discusses related work, followed by section on notations and the formal problem definition. Section~\ref{sec-maximizingProfit}  derives the expected profits and proposes the optimization method. Subsequently, we discuss running time minimizations. Next section presents extensions of the problem to deal selection, admissibility, and bidding of traditional ads. Section~\ref{sec-evaluations} present the experimental evaluations and results. Finally we present our conclusions in Section~\ref{sec-conclusions}.

\section{Related Work}
\label{sec-relatedWork}
Grabchak \emph{et al.}~\cite{grabchak2011adaptive} addressed the problem of optimal selection of guaranteed (group buying) ads . Our work is different, since we deal with optimal bidding, whereas Grabchak \emph{et al.} does not consider the bidding and consider  offline selection of deals. Further, even the non-bidding selection sub-problem discussed in this paper is different since we consider a minimum number of conversions like deals, whereas  Grabchak \emph{et al.} consider an exact number of required conversions.

Different models of group-buying auctions and bidding mechanisms has been studied~\cite{anand2003group,chen2002bidder}. But our problem of bidding to sell deals online---mostly made popular after the emergence of dail-deal sites---has not been studied for any of the group-buying auction models.

Considering related problems of allocation and bidding of display ads, Ghosh~\emph{et al.}~\cite{ghosh2009bidding} considered allocating guaranteed display impressions matching a quality distribution representative of the market. Vee~\emph{et al.}~\cite{vee2010optimal} analyzed the problem of optimal online matching with access to random future samples. Boutilier~\emph{et al.}~\cite{boutilier2008expressive} introduced an auction mechanism for real time bidding of display ads.

There are a number of papers on optimal ranking of textual ads in presence of budget limits. Mehta~\emph{et al.}~\cite{mehta2007adwords} deal with the problem of optimal allocation of textual ads considering budget limits of the advertisers. Buchbindar~\emph{et al.}~\cite{buchbinder2007online} provided a simpler
primal-dual based analysis achieving the same competitive ratio. These papers consider ranking/allocation of textual ads than deals. Further these  problems have  an upper limit on number of impressions, rather than a lower limit as in our problem. Hence, unlike these problems,  ours is not a generalized online bipartite matching.

 %The related area of dynamic mechanisms deals with highly unpredictable environments---like auction agents entering at different times; and optimal allocations are often intractable~\cite{gallien2006dynamic}.
 With the increase of ad exchanges offering real-time bidding, there are a few papers on related problems. Chen~\emph{et al.}~\cite{chen2011real} formulated the problem of supply side allocation of traditional ads with upper bounds on budgets as an online constrained optimization matching problem. Chakraborty~\emph{et al.}~\cite{chakraborty2010selective} considered the problem of ad exchanges calling out a subset of ad-networks without exceeding capacity of individual networks for real time bidding. To the best of our knowledge, the optimal bidding problem of group-buy deals and the extensions have not been addressed.

\section{Notations and Problem Definition}
\label{sec-notaionsDefinition}
 Every group-buy deal  $g$ has a  required minimum clicks  $m$, an expiry time $e$, a cost per click (CPC) $\rho$, and a click through rate (CTR) $\mu$. Thus a deal may be represented as,
\begin{displaymath}g = \langle   m, e, \rho, \mu  \rangle  \label{eqn-dealDefinition} \end{displaymath}

For the rest of the paper our discussions are based on guaranteed number of clicks for the ease of description. The discussions and results are equally applicable for  guaranteed conversions (refer to Section~\ref{sec-gnerelizations} for the details) (by substituting conversion rates (CVR) for click through rates and click per action (CPA) for CPC) and guaranteed displays (by setting click through rate to one and substituting Cost Per Impression (CPI) for CTR).

Let  $\psi_t$ be a binary indicator variable, with $\psi_t=1$ if the advertiser's bid is successful at time $t$---i.e. he wins the bid for impressions and pays the content owner---and zero otherwise.  Let $c_t$ be the number of clicks at time $t$. For our discussions, the time $t$ denotes $t$ user visits (impression opportunities) rather than wall clock time. For a deal $g$ the profit $\mathcal{P}_t$ at time $t$ is,
\begin{equation}
\mathcal{P}_{t} = \left\{     \begin{array}{c c c}
                    \rho c_t - \sum_{j=0}^{t}h(b_j)\psi_j &  \mbox{ when } &  c_t \ge m  \\
                   - \sum_{j=0}^{t}h(b_j)\psi_j        &  \mbox{ when }  & c_t < m
               \end{array}\right.
\label{eqn-profitForumuation}
 \end{equation}
where $h(b)$ is a mapping from bids to the payment whose closed form depends on the auction model, the number of other bidders and the bid distributions. For the commonly
used first price auction for display ads $h(b_j)= b_j$. For other auctions like second price auction, closed forms can be derived based on order statistics~\cite{easley2010networks}. After fulfilling guarantees (i.e. $c_t \ge m$) the expected profit function for the guaranteed deals are the same as that of the traditional non-guaranteed ads. Hence, the period of interest for our analysis and experiments is the time before guarantees are fulfilled.

To maximize the  profit in Equation~\ref{eqn-profitForumuation}, the only parameter decided by the bidder  is the bid amount. Hence we may state the profit maximization problem as,

\textbf{Bidding Problem: } \emph{Given a guaranteed ad $g = \langle  m, e, \rho, \mu  \rangle $, and number of  received conversions $c_t$, find the  bid amount $b_t$ such that the expected profit from $u_t$ user visits is maximal, where $u_t$ is the expected number of user visits before the ad expiry time $e$.}

To explain the nature of the  problem, we start by finding the optimal bid based on the expected values of  parameters at $t=0$. This is the best possible estimate at that point of time. As  time progresses, we will get better estimates of parameters based on the actual values of number of conversions, and user visits so far. Hence we keep updating the optimal bid $b_t$ based on the current state and expected numbers in the future. We assume that $u_t$ is known, as it can be generally estimated from the traffic statistics~\cite{grabchak2011adaptive}.

\section{Maximizing Profit}
\label{sec-maximizingProfit}
We derive the expected profits of group-buy deal campaigns based on the current state of the deal. Subsequently we analyze the nature of the the profit-function,  and present a method to maximize the profits in real-time by bid adjustments.
\subsection{Expected Profits}
\label{sec-expectedProfits}

The  click probability of a deal  is,
 \begin{displaymath}P(click)=P(click|impression) P(impression|bid)\end{displaymath}
The first factor $P(click|impression)$ is equal to the CTR of the deal---is a constant for static auctions considered here.
 The second factor---probability of winning impression $P(impression|bid)$---is an increasing function of the bid amount.
  This implies that the probability of satisfying click guarantees, and consequently the expected profit increase with the bid amount.  On the contrary, the amount paid by the bidder to the publisher ($h(b)$) is an increasing function of the bid amount. Hence  the profit tends to decrease with increasing bid amount. The bids need to be optimized  considering these two conflicting effects on the profit.

 For real-time bidding, different advertisers or intermediaries place bids for a given ad impression. Generally the highest bidder wins, and will display his ad\footnote{Alternatively, bidder with the highest value for bid times CTR may win. This can be easily incorporated into the winning probabilities.}. In general bid values of a bidder varies, either due to the bidder's private value distribution, or due to a deliberate randomization done by the bidder to avoid giving advantage to the competition~\cite{ghosh2009bidding}. Hence, the event of winning is probabilistic, with a binary outcome. Further, winning in consecutive bids can be assumed to be independent of each other.   Hence bidding to win impressions are Bernoulli trials with success probability increasing with the bid amount.

   The users click with probabilities equal to the estimated CTR of the winning ad. This is again a Bernoulli trial with success probability equal to  the CTR. Hence these two trials---bidding and getting conversions---may be combined as a single Bernoulli trial of bidding to win clicks. The probability of success for this composite trial is equal to the product of CTR and probability of winning an impression.
 %Further if we exclude the requirement of conversions (by setting CTRs to one) the trials becomes bidding for guaranteed number of impressions just. Alternatively, if we replace the CTR by CVR, the trial becomes bidding for Groupon Style deals. Thus the analysis and results below are equally applicable for bidding for guaranteed conversions, conversions and displays.\Comment{Raju}{May be this conversion, impression discussion is lot of repetition in the paper, look back after completing generalization section}

 For composite Bernoulli  trial described above, the number of successes follows a binomial distribution. To facilitate representing such a binomial distribution, we introduce the following two functions,
\begin{equation}
\Phi(r_t,u_t,b_t, \mu) = \sum_{j=r_t}^{u_t}  \left(\begin{array}{c} u_t  \\ j \end{array}\right) (\mu d(b_t))^j (1-\mu d(b_t))^{u_t - j} \label{eqn-phiExpression}  \end{equation}
\begin{equation}
\Theta(r_t,u_t,b_t, \mu) = \sum_{j=r_t}^{u_t} j \left(\begin{array}{c} u_t  \\ j \end{array}\right) (\mu d(b_t))^j (1-\mu d(b_t))^{u_t - j} \label{eqn-thetaExpression}
\end{equation}
where $\mu$ is the CTR of the ad, and $r_t$ is the additional number of clicks required to satisfy the guarantees.

Function $d(b)$ is a mapping from the bid value to the probability of winning the impression. For a sealed bid auction in which the highest bid wins (e.g. first or second price auctions), this probability is $d(b)=CDF(b)^{n-1}$, where $CDF$ is the cumulative probability distribution of the bids of other bidders, and $n$ is the total number of bidders. To get a closed form of $d(b)$ we need to assume a
distribution function of bids. For example, if the bids are uniformly distributed between $l$ and $u$, $d(b) = \big((b-l)/(u-l)\big)^{n-1}$. Similar closed forms can be derived for other distributions, and even for cases where different competitors following different distributions~\cite{easley2010networks}.

At optimal profits, the bids are the best responses to competitors and hence are in a symmetric Bayesian-Nash equilibrium~\cite{easley2010networks,krishna2009auction}. Consequently, we may limit our analysis to truthful bidding without the loss of generality as stated by the revelation principle~\cite{dasgupta1979implementation}. Hence the assumptions on bid distributions  above are equivalent to the same assumptions  on private value distributions of bidders at the optimal profit outcomes.

Now the net expected profits is given by the objective
 function,
\begin{eqnarray}
E(\mathcal{P}_{t}) & = &  c_{t}\rho  \Phi(r_t,u_t,b_t,\mu) + \nonumber \\
&&\!\!\!\!\!\!\!\!\! \rho \Theta(r_t,u_t,b_t,\mu) \left(\sum_{j=1}^{t-1}\psi_j h(b_j) + u_t d(b_t)h(b_t) \right)\label{eqn-expectedProfitSingleDeal}
\end{eqnarray}
Please refer to Appendix~\ref{appendix-secExpectedProfitDerivation} for the derivation of the expected profits.

\subsection{Optimizing Expected Profits}
\label{sec-optimizingBids}
 The expected profit in Equation~\ref{eqn-expectedProfitSingleDeal} has to be optimized with respect to the bid amount. An option is to differentiate the function with respect
 to $b_t$ and solve the derivative for zero. But this is hard since the derivative may have large number of terms, and solving the derivative will be harder than a direct approach. Hence a direct optimization of the objective function---as we do below---is  faster.

\begin{figure}[t]
\centering
\includegraphics[height=55mm,width=70mm, trim=25mm 67mm 35mm 67mm]{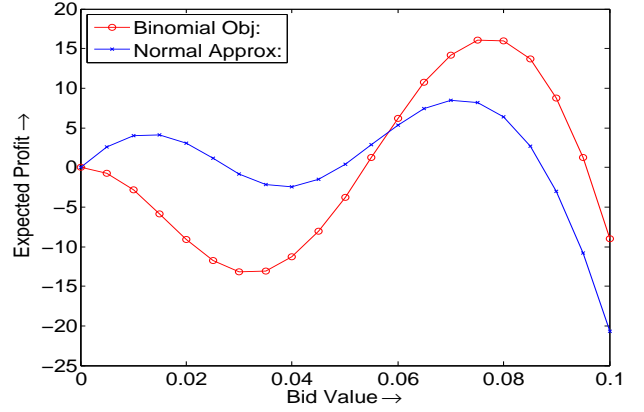}
\caption{Variation of objective function against bid, $c_t=20$, $r_t=5$, $\mu=0.002$, $\sum \psi_i = 0$, and single competitor with uniform random bid in $[0,0.1]$ (i) Exact binomial for $u_t=3000$, $\rho=15$. (ii) Normal approximation for $u_t=2500$, $\rho=13$.}
\label{fig-objectiveNonConvex}
\end{figure}

 An example curve of variation of the objective function with bids is shown in Figure~\ref{fig-objectiveNonConvex}. Two observations significant to the numerical optimization  are (i) the optimization  is non-convex. (ii) the function is not even quasi-convex (unimodal). This implies that a bisection or gradient descent method may get trapped in a local optimum, and hence the convergence to the global optima is not guaranteed. Consequently, we need to start the optimization from multiple points making the problem harder.

 For the bidding process, the  winning probability is one if the bid is greater than the maximum bid of the competitors bid distribution; and zero for bids less than the minimum bids. Hence the optimal values will always be between the maximum and minimum even without imposing external constraints. This allows a simpler unconstrained optimization. The optimizer restarts the search from multiple random starting points to avoid local minima traps (the details of the restarts are discussed in the Section~\ref{subsec-minimizingrunning}). Further since the optimization is online, fast-convergence is highly desirable. Considering these factors, we adapt Brent's optimization method. Brent's optimization combines parabolic interpolation with golden ratio search for faster convergence. If the parabolic interpolation fails, the search falls back to the golden ratio search.

\section{Running Time Minimization}
\label{subsec-minimizingrunning}
The optimization of the bids has to be performed online between the impression opportunities. Evaluating objective function in Equation~\ref{eqn-expectedProfitSingleDeal} directly may involve computing hundreds of thousands of terms. Since the time duration between the impression opportunities can be very small, the optimization time must be within acceptable  limits. In addition to using  fast converging Brent's optimizer, we adopt several  approximations and computational methods for improved running time.

\textbf{Reducing Number of Terms: }
Typically for deals  the user visits needed to get the required number of clicks are very large compared to the clicks needed for tipping the deal. i.e.  $u_t\gg r_t$; except for a last few user visits before deal expiry. Exploiting this, we reduce the computation time by rewriting the Equations \ref{eqn-phiExpression} and \ref{eqn-thetaExpression} as,
\begin{equation}
\Phi(r_t,u_t,b_t, \mu) = 1- \sum_{j=0}^{r_t-1}  \left(\begin{array}{c} u_t  \\ j \end{array}\right) (\mu d(b_t))^j (1-\mu d(b_t))^{u_t - j} \label{eqn-phiExpressionTuned}  \end{equation}
\begin{eqnarray}
\Theta(r_t,u_t,b_t, \mu) & = & u_t\mu d(b_t) - \nonumber \\
&&\!\!\!\!\!\!\!\!\!\!\!\!\!\!\!\!\!\sum_{j=0}^{r_t-1} j \left(\begin{array}{c} u_t  \\ j \end{array}\right) (\mu d(b_t))^j (1-\mu d(b_t))^{u_t - j} \label{eqn-thetaExpressionTuned}
\end{eqnarray}
This rewriting may reduce computations from  hundreds of thousands of terms to less than a few hundred terms.

\textbf{Normal Approximation: }In spite of replacing $u_t$ by $r_t$, computing binomial CDFs in Equation \ref{eqn-phiExpressionTuned} and \ref{eqn-thetaExpressionTuned} may involve summation of hundreds of combinatorial terms. The binomial CDF may be approximated by a single term normal CDF for large values of $u_t$ (central limit theorem). Exploiting this, we compute $\Phi$ and $\Theta$ based on a normal CDF with correction for continuity. i.e.
\begin{eqnarray}
\Phi(r_t, u_t, b_t, \mu) &\approx & 1-  \frac{1}{\sqrt{2\pi u_t\mu d(b_t) (1-\mu d(b_t))}} \times \nonumber \\
&&\int_{0}^{\frac{r_t-0.5- u_t\mu d(b_t)}{u_t\mu d(b_t)(1-\mu d(b_t))} }e^{-\frac{(x-u_t\mu d(b_t))^2}{2 u_t\mu d(b_t) (1-\mu d(b_t))}} dx \nonumber
\end{eqnarray}
There is no analytical solution for this integral, but can be looked up from a normal CDF table or can be approximated by finite analytical forms.

Similarly, approximating the standardized form of $\Theta$ as $\Theta^\prime$,
\begin{eqnarray}
\Theta^\prime(r_t, u_t, b_t, \mu)  & = & \frac{1}{\sqrt{2\pi}} \int_{\frac{r_t-0.5-d(b_t) u_t}{\sqrt{u_t\mu d(b_t)(1- \mu d(b_t))}} }^{\infty} z e^{\frac{-z^2}{2}} dz \nonumber \\
 &=&  \frac{1}{\sqrt{2\pi}}e^{-\frac{(r_t-0.5-u_t\mu d(b_t))^2}{2 u_t\mu d(b_t) (1-\mu d(b_t))}} \nonumber
\end{eqnarray}
where $z=\frac{(j-u_t\mu d(b_t))}{\sqrt{u_t\mu d(b_t) (1-\mu d(b_t))}}$

\begin{eqnarray}\Theta(r_t, u_t, b_t, \mu) & \approx  & \sqrt{u_t\mu d(b_t) (1-\mu d(b_t))}  \Theta^\prime(r_t, u_t, b_t, \mu)  + \nonumber \\
&&  u_t\mu d(b_t) \nonumber
 \end{eqnarray}
 For small  $u_t$ and $r_t$,  normal approximation may diverge more from the original binomial function, and computation of the binomial is less costly. Hence it is more accurate to use actual binomial distribution for smaller $r_t$ and $u_t$.  We depend on the common rule of thumb for approximating binomial CDF by normal CDF, i.e. if $u_t\mu d(b_t)(1-\mu d(b_t)) \ge 10$ we use normal approximation.

Considering the optimization of the normal approximation, the sample graph of the objective function with the approximations is shown in Figure~\ref{fig-objectiveNonConvex}. Like the original binomial objective, the normal approximation is neither convex nor quasi-convex. Consequently optimizing the approximation faces the same difficulties as the optimization of the original objective.

\textbf{Setting the Starting Points: } The optimal bid values generally change only nominally for subsequent impressions of a deal. Exploiting this fact, the optimal bid for an impression is used as the starting point for optimization for the next impression. This reusing of optimal bids expedite convergence.

\textbf{Multiple Starting Points: } The non-convexity  of the objective requires the optimal value search to start from multiple points. As optimal bids change only incrementally for successive impressions of a deal, we avoid restart from multiple points for every impression. For this---instead of starting from previous optimal values as described above---we chose random starting points for optimization for the first twenty impressions. The bid corresponding to the maximum objective value among these searches is used as the optimal value. This strategy is found to be converging to optimal values for all the deals we tested.

\textbf{Re-Computation Frequency: }  As the change in optimal bids  are nominal for subsequent impression opportunities, the previous bids can be reused. Hence we re-optimize bids only after a number of impression opportunities (every thirty two impression opportunities in our experiments). Further, optimal value is always recomputed if there is a click, since one more clicks may cause a non-trivial changes in the optimal bid.

%
%\textbf{Incremental Computations: }The optimization is incremental---i.e. an approximate value is available at every point of time, which is continuously refined in every iteration. Hence we may use the values available at that time if we run out of time before the complete convergence.
%

\section{Extensions to \\Related Problems}
\label{sec-gnerelizations}
\begin{figure}[t]
\centering
\includegraphics[height=55mm,width=77mm, trim=34mm 88mm 35mm 70mm]{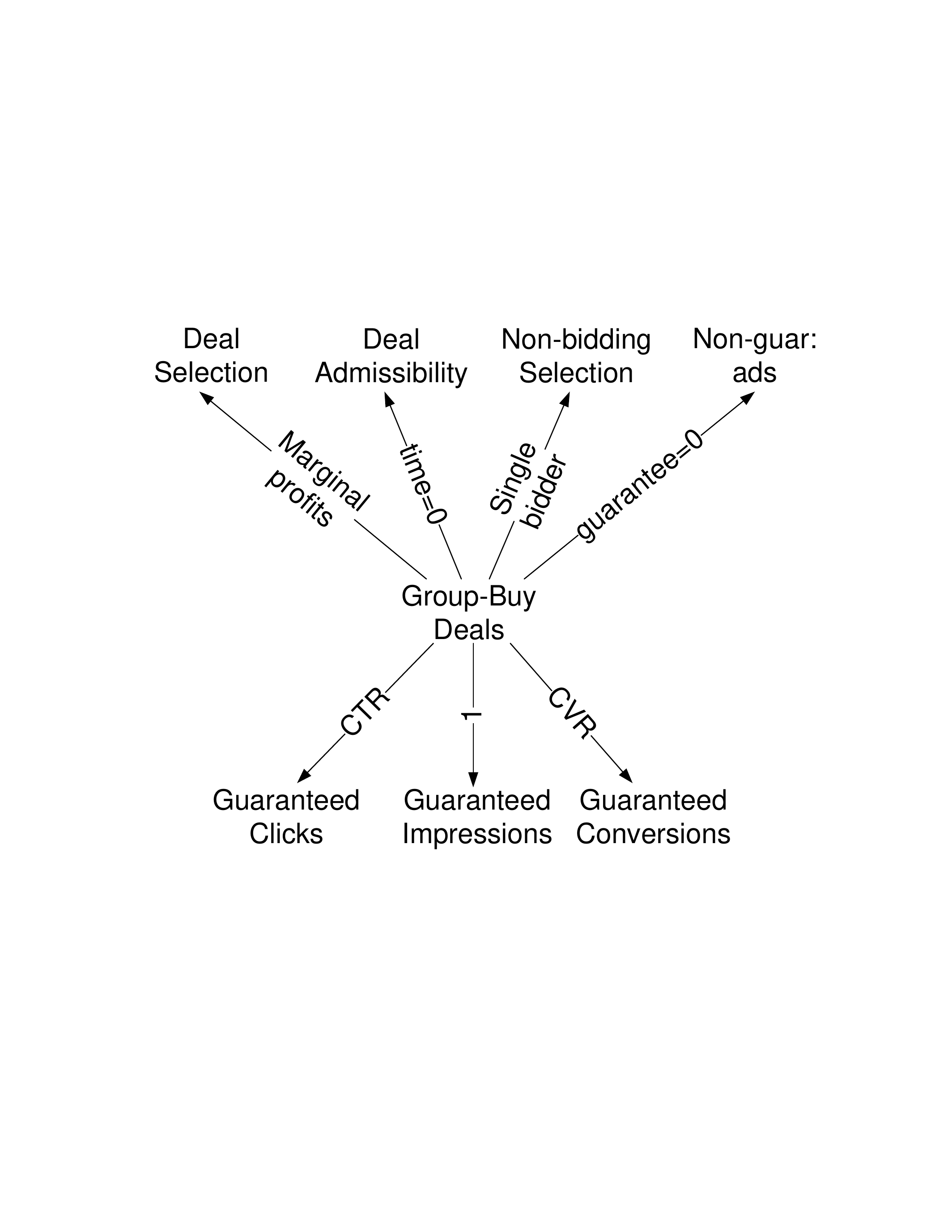}
\caption{Group-buying deal bidding extends to many related problems on specific assumptions. The guarantees can be in terms of conversions, impressions, or conversions depending on the event probabilities used (i.e. CTR, 1 and CVR). Further the objective reduces to deal selection, deal admissibility, selection of deals with no bidding, and to the optimal bidding of traditional non-guaranteed ads on various assumptions on parameters.}
\label{fig-objecvieReductions}
\end{figure}

The optimal bidding for  group-by deals is a general problem, which on specific assumptions reduces to number of related use-cases. As a corollary, the proposed solution  reduces to solutions of these problems on the same assumptions. In Figure~\ref{fig-objecvieReductions} we enumerate these extensions and assumptions on which the guaranteed  bidding will reduce to these problems.

Considering three downward branches in Figure~\ref{fig-objecvieReductions}, the guaranteed bidding easily transform to guaranteed clicks, displays and conversions. For example by setting $\mu=CTR$ the deal definition in Equation~\ref{eqn-dealDefinition} will define a guaranteed conversion campaign. Consequently, on the same assumption the profit  formulation in Equation~\ref{eqn-expectedProfitSingleDeal} gives guaranteed click campaign profits. Similarly, by setting $\mu$ equal to   1 and CVR, the problem and the solution will be applicable for  guaranteed impressions and conversions deals respectively.

Four upward branches in Figure~\ref{fig-objecvieReductions} illustrate reductions to related bidding and selection problems. We discuss  them in the sections below and  the the deal related extensions are evaluated separately in Sections~\ref{subsec-selectingDealsExpt} and ~\ref{subsec-admissibilityDealsExpt}.

\subsection{Deal Selection}
\label{subsec-guarenteedAdSelection}
Deal selection is the problem of maximizing expected profits by choosing the best deal to bid for every impression. Considering the online nature of the problem, we perform a greedy deal selection. The deal with the maximum marginal profit  by the impression is selected as the winner. Expected marginal profit is calculated as the difference between the  expected profits of  winning the impression and failing to win the impression. Adding the bid amount to marginal profit to derive the marginal revenue as,

\begin{eqnarray}
E(v_{it}) &=& \!\!\!\!\! E(\mathcal{P}_{i(t+1)}|\psi_t = 1)- E(\mathcal{P}_{i(t+1)}|\psi_t=0) + b_t \label{eqn-expectedValueIncrease} \\
 & = & \!\!\!\!\! \mu \rho_i \left[(c_t+ r_t -1) \left(\begin{array}{c} u_t-1  \\ r_t -1 \end{array}\right) (\mu d(b_t))^{r_t-1}\right. \nonumber \\
&& \!\!\!\!\!\!\!\!(1-\mu d(b_t))^{(u_t-r_t)}+  \Phi( r_t-1, u_t-1, b_t, \mu) \Big]   \label{eqn-expectedTrueValue}
 \end{eqnarray}
Derivation of the equation is given in Appendix~\ref{appendix-secPrivateValueDerivation}. This marginal revenue is  the expected private value of the impression for the deal bidder, isomorphic to the private value of the traditional ads. Similar to the traditional ads, $E(v_{it})-b_t$ gives the expected profit by displaying deal $i$ at time $t$.

To select a deal, the  bid values of deals are optimized against the expected profits as described in Section~\ref{sec-maximizingProfit}.  These optimal bids are substituted in Equation~\ref{eqn-expectedTrueValue} to calculate the private values. The deal with the highest increase in the expected profit is selected for bidding. We evaluate the proposed selection in Section~\ref{subsec-selectingDealsExpt}. Selection considering deals groups is a harder combinatorial optimization problem, and we leave this for  future research.

\subsection{Deal Admissibility}
\label{subsec-admssibility}
Admissibility criterion decides if the profit from a deal is likely to be positive.  Admissibility  can be directly derived as a special case of optimal bidding. More specifically, if the maximal expected profit from an deal is positive at  $t=0$ the ad is admissible. At $t=0$ the Equation~\ref{eqn-expectedProfitSingleDeal} reduces to,
\begin{equation}
E(\mathcal{P}_{t})  =   c_{t}\rho  \Phi(r_t,e,b_t,\mu) + \rho \Theta(r_t,e,b_t,\mu) - e d(b_t)h(b_t)
\label{eqn-admissibility}
\end{equation}
where $e$ is the expected number of visits before the deal expiry. To evaluate admissibility, bids are optimized for maximal profits  and substituted in Equation~\ref{eqn-admissibility}. If the expected profits are positive, we consider the deal as admissible\footnote{Instead of zero,  a positive profit threshold may be used as the admissibility criterion.}. The profit increase from admissibility is evaluated in Section~\ref{subsec-admissibilityDealsExpt}.
\subsection{Non-Bidding Selection}
 If there are no competing bidders---i.e. number of bidders is equal to one---the problem will reduce to that of selection from competing deals. Alternatively, this scenario may be thought of as the publisher selecting deals directly. In this case $d(b_t)=1$ in selection Equation~\ref{eqn-expectedTrueValue} for all values of $b_t$. Since this is a special case of the selection with bidding, we do not include separate evaluations, as results are directly implied. Alternatively, since there is no bidding involved, selection may be modeled as an offline problem. In this case, selection of deal sets for maximization profits of deal combinations may be of interest. We leave this optimal offline selection of deal sets for  future research.%\footnote{Grabchack \emph{et al.}~\citeyear{grabchak2011adaptive} derived solutions for a related problem from greedy solutions of stochastic-knapsack.}

\subsection{Non-Guaranteed Ads}
\label{subsec-NGads}
As shown in Figure~\ref{fig-objecvieReductions}, when the minimum click guarantee is zero, the guaranteed deal bidding  reduces to the bidding of tradition non-guaranteed ads. Consequently, the expected profits for non-guaranteed ads may obtained by substituting $r_t=0$ in Equation~\ref{eqn-expectedProfitSingleDeal} as,
\begin{displaymath}
E(\mathcal{P}_{t})  =  c_{t}\rho   + \rho \mu u_t d(b_t) - \left(\sum_{j=1}^{t-1}\psi_j h(b_j) + u_t
d(b_t)h(b_t) \right)\label{eqn-expectedProfitNonGuaranteed}
\end{displaymath}
Considering the profit from a single impression (i.e. $u_t=1$), thereby ignoring the constants terms of past profits, the expected profit becomes, $E(\mathcal{P}_{t})  =   d(b_t)(\rho \mu   - h(b_t))$. This expected profits can be maximized with respect to the bid values. For example, for a first price auction with $n$ bidders and  competitors having a bid distribution  with cumulative distribution function of $F(v)$, $d(b_t) = F(b_t)^{n-1}$ and $E(\mathcal{P}_{t})  =   F(b_t)^{n-1}(\rho\mu  - b_t)$.

Note that the expected profits, hence the optimal bids, derived above is the same as the existing formulations of optimal static bid profits of traditional ads~\cite{easley2010networks,krishna2009auction}. This is a manifestation of the broader fact that the existing bidding formulations are optimal for static environments.  Hence the optimal real time bids will essentially be the same. In other words, the  real-time bidding provides no advantage over static bidding for  non-guaranteed ads in static environments.  On the other hand, the real-time bidding would improve profits over existing bidding for dynamic environments~\cite{chen2011real}. Thus the proposed bidding provides a strategy to account for the dynamism in parameters of traditional ads---like the changed estimates of click-through rate or competitor bid distributions.

\begin{figure*}[ht]
\centering
\subfigure[]{\label{fig-biddingProfitBasic-a}
\includegraphics[width=80mm,height=65mm,trim=28mm 67mm 26mm 75mm]{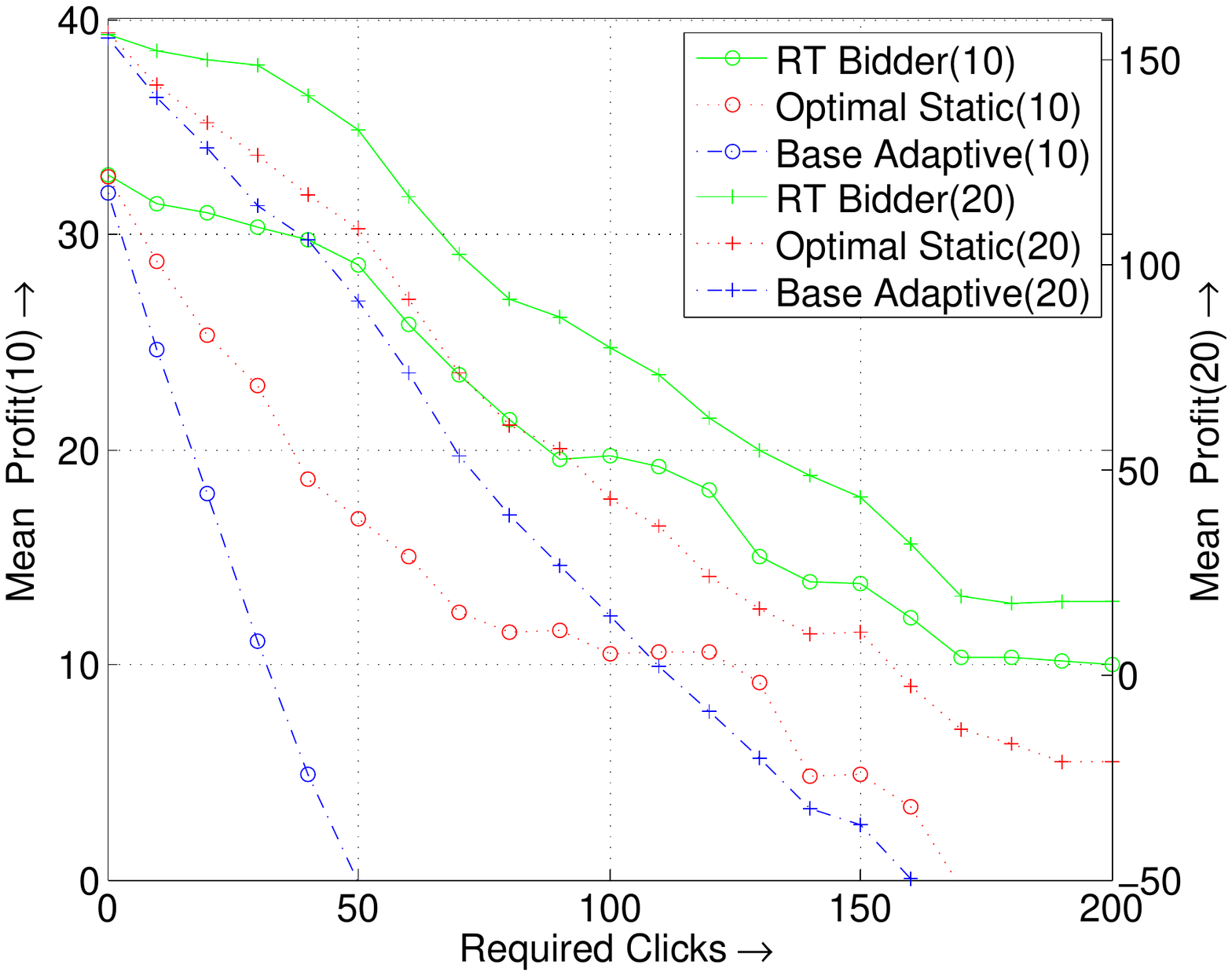}
}
\hspace{0.5cm}
\subfigure[]{\label{fig-biddingProfitBasic-b}
\includegraphics[width=80mm,height=65mm,trim=21mm 66mm 29mm 73mm]{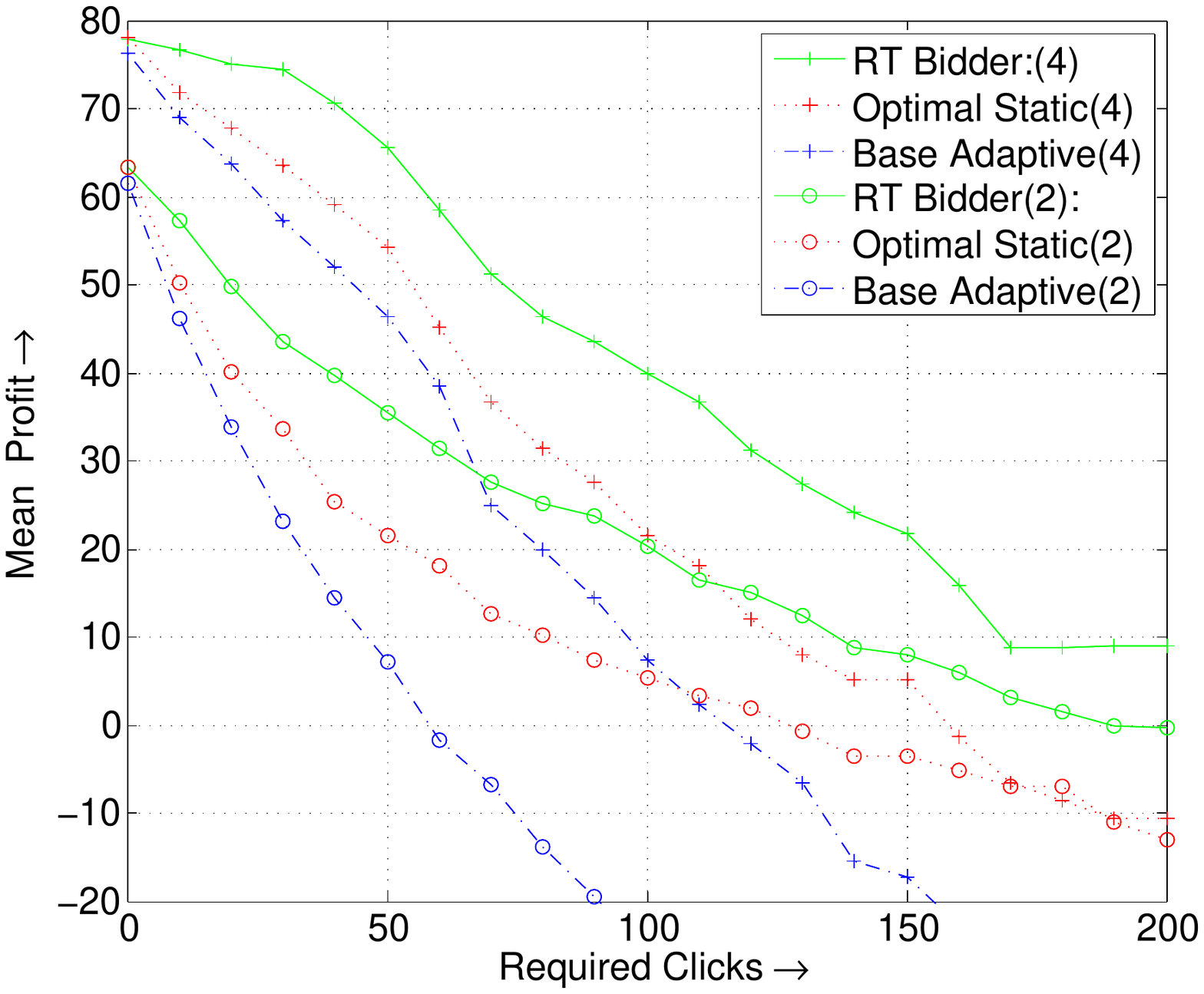}
}
\caption{Comparison of mean profits of RT bidder with optimal static bidder and the baseline adaptive bidder. (a)  Four bidders, competing bid distributions are uniform random in $[0, 0.4]$, pay per click ($\rho$) is 10. For  group~(20)   $\rho$ is doubled to 20, keeping other parameters the same. (b) Group~(4) has four bidders, with  competing bids uniform in $[0,0.2]$, keeping the other parameters same as group~(10) in Figure \ref{fig-biddingProfitBasic-a}.  Group~(2) has two bidders instead of four in group~(10).
}
\label{fig-biddingProfitBasic}
\end{figure*}
\section{Evaluations and Results}
\label{sec-evaluations}
We compared the profits by the proposed bidding strategy with baseline strategies of (i) optimal static bidding (ii) a basic real time bidding. We evaluated the profit increase, running time,  deal selection and admissibility---including the robustness of our method to violation of assumptions.

\mund{Data Set: } From a click log of 330 million impressions, we randomly selected 935 ads (with 9.3 million impressions) having a minimum of 5 clicks and 1000 impressions. Click log contains impressions  and  whether the impression resulted in a click or not.  %The click logs were not of deals, but traditional ads. Hence we had to assign guarantees and expiry timings for different experiments as described below. Since the bid values are not registered in the click log, we had to draw competitor bid values from  assumed distributions randomly.

\mund{Baseline Bidders: }First baseline bidder is an optimal static bidder. The bidder derives optimal bids as a function of number of competitors and their bid distributions~\cite{easley2010networks}, and is optimal if there are no guarantees. The second baseline bidder is a basic adaptive bidder, which bids as $staticOptimal+r_t/u_t -CTR$. The strategy is intuitive, as it increases bid over the static optimal bid if the required click rate is greater than the CTR and decreases the bids otherwise. We also used the profits by a placing a random bid as a baseline in our initial experiments. Random bidding performed much worse than all of the above baselines and is not plotted in the results.

\subsection{Bidding Profit Comparison}
 To compare the profits by bidders, the proposed real-time (RT) and baseline bidders  compete with random bidders for every ad in the ``replayed"  click log. For example, the RT bidder places its bid for the first impression of an ad, along with the competing random bidders. If the RT bidders' bid is the highest, the bidder wins the impression. The deal wins a click if the click log indicates a click for that impression. Then the same  process repeats for the second impression and so on. Similarly, other baseline bidders are made to compete with the same random bidders for the same set of ads, and the realized profits are compared. This replaying reproduces online experiment, since the user action in the consecutive impressions are most likely from different users, and hence independent. The experiments are repeated by changing every significant parameter---one at a time---to analyze the effects of different parameters.

 Since the click logs are of traditional ads having no required clicks or expiry, we set the expiry timings as the number of impressions of the ad in the log. To compensate for using the traditional ads instead of deals, the required clicks are varied by a parameter sweep  over the  plausible range. We do not vary expiry time, since the ratio of required clicks to the expiry time determines the profit---rather than the expiry time alone---and this ratio varies as we do a parameter sweep on required clicks. Further, note that different ads in the set of 935 ads have different expiry timings, effectively functioning as a parameter sweep on expiry times. CTR of every ad is estimated as the ratio of number of clicks to total impressions\footnote{CTR estimation and prediction is a separate problem researched extensively~\cite{richardson2007predicting}.}.

  The competing bids are selected from different random distributions (bids are not registered in the click log), since the bidders generally randomize their bids to avoid the competing bidders guessing their bids~\cite{ghosh2009bidding}. Note that this randomization in bids may be achieved by randomly choosing different deals to bid for in different time slots even for the proposed optimal bidding strategy. At the Bayesian-Nash equilibrium---in which everyone follow the optimal strategies---these assumptions on bid value distributions are equivalent to the same assumptions on private values of the advertisers, as mentioned in Section~\ref{sec-expectedProfits}. Further, the maximum entropy (i.e. minimum assumptions) uniform random competing bid distribution  is the hardest to predict and to optimize against (hence we use this distribution for the experiments below). Any other distributions, including a fixed optimized competing bid is easier to optimize against and the realized profits will be higher.

 %These experiments are not a complete substitute for an online evaluation since they do not consider the dependence of conversions on other ads displayed in between. But the use of real conversion logs account for the burstiness of conversions in the actual click streams.

Figures~\ref{fig-biddingProfitBasic-a} and ~\ref{fig-biddingProfitBasic-b} show four  sets of comparisons of profits realized by different bidders against required minimum clicks $m$.
In the first set of Group~(10) in Figure~\ref{fig-biddingProfitBasic-a}, there are four bidders---the bidder evaluated and three other random bidders. Random bidders bid in a uniform random distribution in the interval of $[0, 0.04]$.   Pay per click ($\rho$) is set to \$10 for this group. The remaining groups of experiments in Figures~\ref{fig-biddingProfitBasic-a} and \ref{fig-biddingProfitBasic-b} are designed to analyze the effects of changes in parameters.

\begin{figure}[t]
\centering
\label{fig-biddingPoriftRobustness}
\includegraphics[width=77mm,height=60mm,trim=20mm 68mm 21mm 75mm]{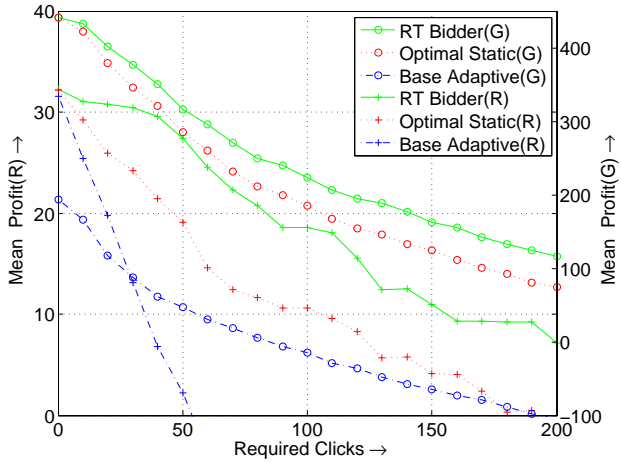}
\caption{$Group (G):$ Normally  distributed competing bid distributions with $mean=0.02$ and $\sigma=0.01$. All other parameters are same as Group~(10) in Figure~\ref{fig-biddingProfitBasic-a}.  $Group (R):$ Robustness to deviation from the assumed distributions.   Two of the four competitors' bids are normally distributed with $mean= 0.02$ and $\sigma=0.01$  instead of the assumed unform distribution  in $[0, 0.04]$ by the RT and static bidders.
}
\label{fig-normal-robust}
\end{figure}

Analyzing common trends in all these groups, the profit of the real time bidder exceeds that of  the baseline bidders for every $m$ in all the experiments. As expected, the profits reduces with $m$. The increase in profit is as large as 70-150\%  (e.g. at $m=150$ for Group~(10) increase is $(13.74-4.86)/4.86 =1.45$\big).  As an exception, the profits of real time bidder is the same as the static optimal bidder at $m=0$. This is an implication of discussions in Section~\ref{subsec-NGads}, that the current static optimal bidding of non-guaranteed ads are a special case of the proposed  RT bidding. As  $m$ increases from left to right, the absolute and percentage of difference between static and RT profit  increases. The baseline adaptive bidder performs worse than static bidder for most parameter combinations, since the bidding considers only a subset of parameters. This is a manifestation of the fact that a simpler intuitive adaptive strategy is not likely to perform well, especially since optimal bid depends on a large number of parameters.  The baseline adaptive bidder, at $m=0$ perform very similar to (but not the same) as the static optimal bidder for uniform competitor bids (for normal distribution  in Figure~\ref{fig-normal-robust}, they   differs considerably at $m=0$) as typical CTR values are very small compared to the optimal bids.

The next three sets of experiments analyze the effects of each of the three parameters---$\rho$, bid distribution, and number of bidders. Parameters are changed one at a time with respect to Group~(10) in Figure~\ref{fig-biddingProfitBasic-a} by a factor two. Figure~\ref{fig-biddingProfitBasic-a} Group~(20) (plotted against the second $y$-axis) increases the profit due to increase in pay per click $\rho$. The increase in profit is more than linear to $\rho$ since the revenue increases linear to $\rho$, but the cost of impression remains the same, consequently the profit ($revenue-cost$) increases many times. For the Group~(4) in Figure~\ref{fig-biddingProfitBasic-b}, reduced bids of the competing bidders result in lower optimal bids hence increase in profits. In the second set of experiments in Figure~\ref{fig-biddingProfitBasic-b} group~(2), the reduced number of competing bidders leads to lower optimal bids and hence increase in profits.

The next sets of experiment in Figure~\ref{fig-normal-robust} further relax assumptions on competing bidders. The  group~(G) has competing bidders having Gaussian bid distributions instead of the uniform random distribution in the previous experiments; and group~(R) evaluates the robustness of RT bidding against violation of assumptions. Like the uniform distribution, for the Gaussian  experiments in group~(G) as well, the RT bidder outperforms the competitors. The profits are higher than the uniform distributions in Figure~\ref{fig-biddingProfitBasic-a} group~(10),  since the lower entropy of Gaussian distribution is easier to optimize against.  For the robustness experiments if group~(R), the RT and static bidders assume uniform distribution in $[0,0.04]$ for the three competing bidders, but two of the competing bidders bid in normal distribution. The RT bidder still dominates over the baseline by considerable margins. A plausible explanation for similar profits to Figure~\ref{fig-biddingProfitBasic-a} Group~(10) is that the effects of violation of assumption and the easier Gaussian distribution  cancel each other.

\begin{figure}[t]
\centering
\includegraphics[width=77mm,height=60mm, trim=40mm 89mm 47mm 87mm]{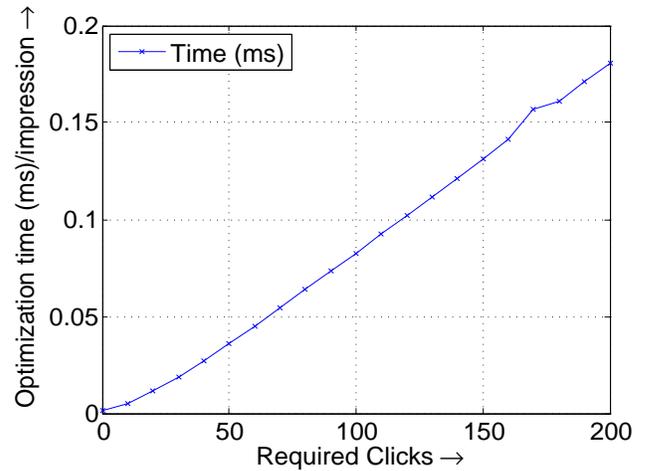}
\caption{Optimization time increases linearly with the required clicks, but is in acceptable limits.}
\label{fig-runningTime}
\end{figure}

\subsection{Running Time Evaluation}
 These experiments are conducted on a shared 16GB RAM machine with two dual core CPUs running at 2.54 GHz. The running time optimizations described in the Section~\ref{subsec-minimizingrunning} are applied (similar to other experiments). Analyzing  the objective function in Equations~\ref{eqn-phiExpressionTuned} and \ref{eqn-thetaExpressionTuned}, the parameter having maximum effect on the running time is the require clicks $m$. The number of terms in the objective increases with $m$, and other parameters will have negligible effect on computation time. Hence we evaluated mean time to optimize the bids for an impression against plausible ranges of $m$. Figure~\ref{fig-runningTime} shows that the running time increases linearly with the required cliks as expected. The highest running time is less than 0.2 milliseconds, which is quite acceptable.

\begin{figure*}[ht]
\centering
\subfigure[]{\label{fig-extensions-a}
\includegraphics[width=77mm,height=60mm,trim=40mm 87mm 50mm 92mm]{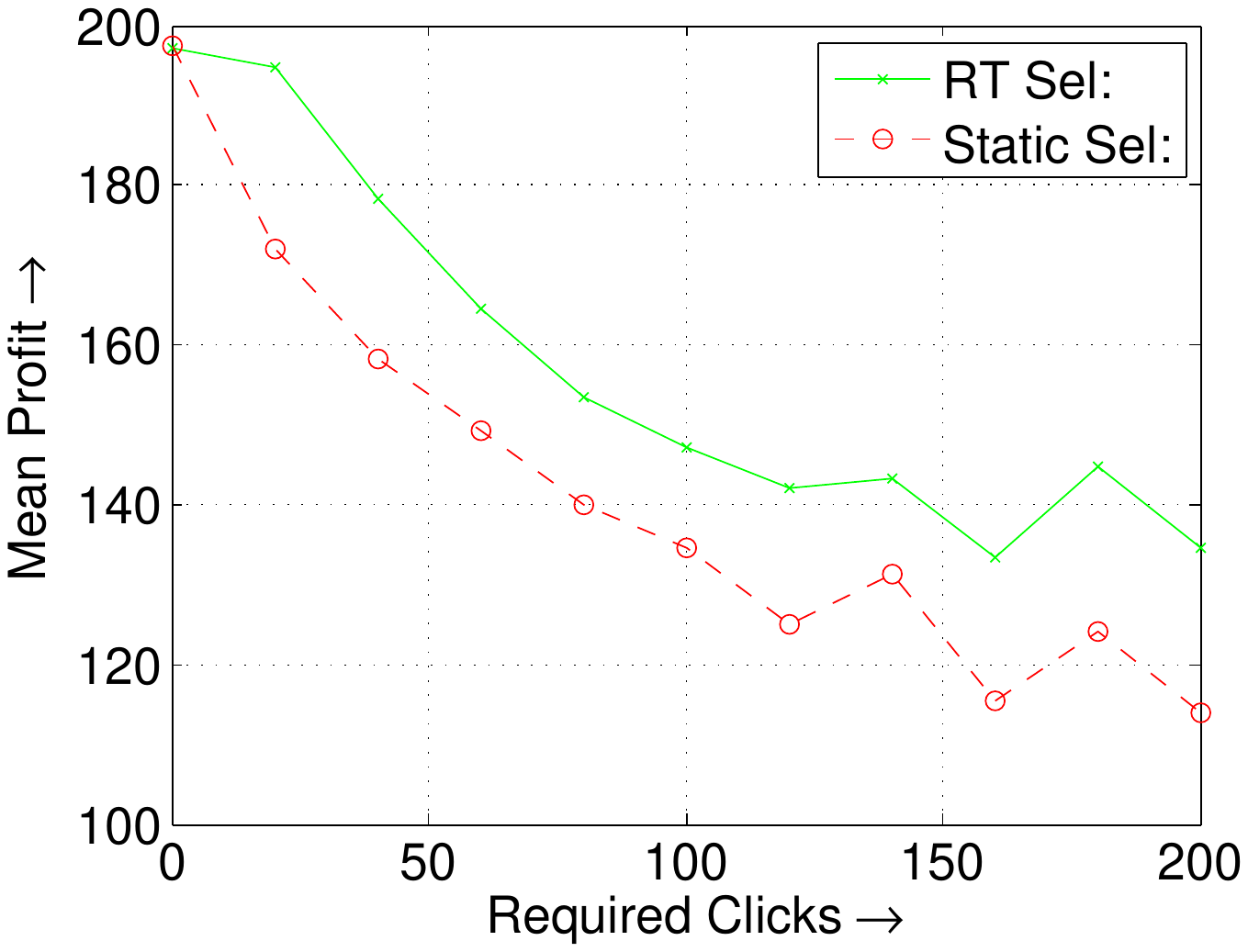}
}
\hspace{0.5cm}
\subfigure[]{\label{fig-extensions-b}
\includegraphics[width=77mm,height=60mm,trim=25mm 73mm 33mm 83mm]{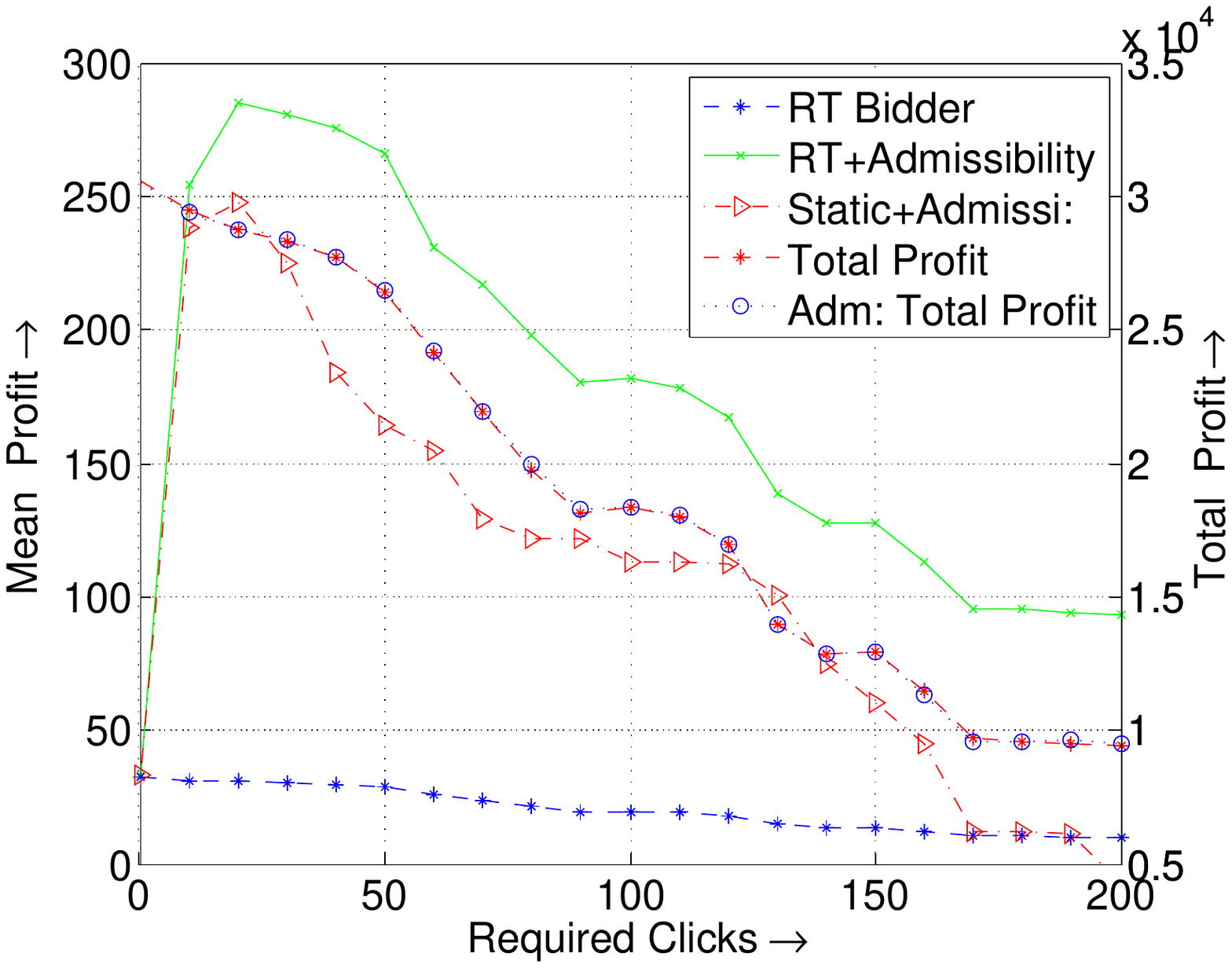}
}
\caption{ Evaluation of extensions: (a) Comparison of real time and static selection.  (b) Admission control increases mean deal profits significantly, for both static and RT bidder. The total  profits (plotted against the second $y$-axis) of RT bidder remains the with and without admissions control, indicating lack of false negatives.}
\label{fig-extensions}
\end{figure*}

\subsection{Deal Selection}
\label{subsec-selectingDealsExpt}
We evaluate the proposed real time deal selection described in Section~\ref{subsec-guarenteedAdSelection}. Selection is compared with the optimal static selection preferring higher private value ads computed as $CTR\times CPC-bid$. Among the 935 ads in click log, we removed ads with CTR greater than 0.02 for these experiments. These high CTR ads make the selection easy by dominating over other ads, hence make it harder to distinguish the selection quality.  Among 823 remaining ads, we randomly created groups of four ads. Selection experiments are performed for each group separately, and mean profits among all groups are plotted.  We set the bidding parameters as $\rho=20$, total number of user visits as 15000, time to expire for each ad as minimum of 15000 and the number of impressions of the ad, number of bidders as four, and competing bidders' bids are uniform random in $[0,0.04]$. For every deal, the require clicks is set uniform random between zero and the maximum value shown in the $x$-axis to have different required clicks values for different deals.

 For selecting the best ad in the group for an impression, first the bids are optimized using the real time bidder. These optimal bids are used in Equation~\ref{eqn-expectedTrueValue} for real time selection, and to compute private value (i.e. $CTR\times CPC-bid$)  of the static selection.  To separate improvement in profit by selection from improvement by bidding, the proposed RT  bidder is used for bidding after both the static and RT selections.  The mean realized profits are shown in Figure~\ref{fig-extensions-a}. When the required clicks is zero, the real time bidding gives the same optimal profit as the static bidding (keep in mind that the static selection is optimal when require clicks is zero). For higher values of required clicks, the real time bidder gives considerably higher profits, with percentage of increase in profit increasing with required clicks. The profit swings at larger values of required clicks is due to the random factors in assignment of required clicks.
\subsection{Deal Admissibility}
\label{subsec-admissibilityDealsExpt}
The admission control proposed in  Section~\ref{subsec-admssibility} is evaluated by comparing mean deal profits against the profits without admission control. Similar to selection, bids are optimized and substituted in Equation~\ref{eqn-admissibility}. The deals giving positive expected profits are passed to the bidder, and mean profits are plotted.  Figure~\ref{fig-extensions-b} shows that admission control improves profits by more than six times for some values of required clicks $m$.  The profit increase for both static and RT bidders, showing effectiveness of admission control independent of the bidding method (like the previous experiments RT bidder performs considerably better than static). At $m=0$ all the ads  in the click log have positive expected profits. The mean profit  no longer decreases monotonically, as the admission control eliminate more low profit ads with increased  $m$. Further the total profits of RT bidder with and without admissibility is almost exactly the same. This shows that there are no false negatives removed. The admission control does not increase the total profits, because for the ads with negative expected profits, the real time bidder will bid zero making the losses to zero. With admissibility, the bidder incur the same profit from much lesser number of ads and user visits, hence he can use the remaining user visits to sell other ads.
\section{Conclusions}
\label{sec-conclusions}
An emerging category of the online ads are the group-buy deals requiring minimum number of purchases. For an advertiser or intermediary  selling these deals, optimizing  bids is necessary for maximal profits. Existing  bidding strategies are sub-optimal for these deals, as they do not consider event minimum group-size guarantees and expiry timings. To this end, we propose a real time bidding strategy for guaranteed deals. We derive the expected profits as a function of the dynamic and static parameters of the deals. These expected profits are shown to be non-convex, and numerically optimized against the bid values. To satisfy the stringent time constraints of online bidding, we use several approximations and running time optimizations.  Exploiting the generality of the proposed formulation, we extend the solution to related problems of deal selection for bidding, admissibility, selection for non-bidding scenarios and real time bidding of non-guaranteed ads. Our empirical comparisons with base adaptive and the existing static strategies on a multi-million click log show significant profit improvements. Further our evaluations show acceptable running time and robustness against the violation of assumptions. Evaluations of extensions show considerable profit improvement by the proposed deal selection and  admissibility.

\bibliographystyle{abbrv}
%\bibliography{optimalbids}

\renewcommand{\theequation}{A-\arabic{equation}}
\renewcommand{\thesubsection}{A-\arabic{subsection}}
% redefine the command that creates the equation no.
\setcounter{equation}{0}  % reset counter
\section*{APPENDIX}
\subsection{Derivation of Equation~\ref{eqn-expectedProfitSingleDeal}}
\label{appendix-secExpectedProfitDerivation} Let $G$ denotes the
event of satisfying the guaranteed number of clicks. Let $R$
and $C$ denote revenues and costs respectively,

\begin{equation}E(\mathcal{P}_{it})  =  P(G)(E(R|G)- E(C|G)) - P(\neg G)( E(C|\neg G) )  \label{eqn-expectedProfitHLevel}\end{equation}

\textbf{Cost: } At time $t$  an amount equal to $\sum_{j=1}^{t-1}\psi_j b_j$ is paid for the impressions.
The future expected cost is the expected payment  till $u_t$.
Let $D$ denotes the total number of displays till $u_t$~\footnote{We compute the expected displays in the event of meeting the guarantees and not meeting the guarantees separately for clarity, since it may look like this evidence influences expectation of displays.},
\begin{eqnarray}
E(D) &= & E(D|G) P(G) + E(D|\neg G)P(\neg G) \nonumber
\end{eqnarray}
These conditional expectations can be expanded as,
\begin{eqnarray}
&=& P(G)\sum_{j=1}^{u_t} j P(D=j|G) + P(\neg G)\sum_{j=0}^{u_t} j P(D=j| \neg G)  \nonumber \\
&=& \sum_{j=1}^{u_t} j \left[P(D=j|G) P(G) + P(\neg G) P(D=j|\neg G) \right] \nonumber \\
&=& \sum_{j=1}^{u_t} j \left[\frac{P((D=j)\wedge G)}{P(G)} P(G) + \right. \nonumber \\
&& \left. P( \neg G) \frac{P((D=j)\wedge \neg G)}{P(\neg G)}\right] \nonumber \\
&=& \sum_{j=1}^{u_t} j \left[P\left((D=j)\wedge G \right) + P\left( (D=j)\wedge \neg G \right)\right] \nonumber \\
&=& \sum_{j=1}^{u_t} j  P(D=j)  \nonumber
\end{eqnarray}
Since number of impressions follows a binomial distribution with success probability equal to probability of display $p_d$, $E(D) =  p_d u_t $. Hence the expected cost is $ = b_t p_d u_t $

\textbf{Revenue: } Revenue is conditional on G, as revue in the
event of $\neg G$ is zero. At time $t$, total expected revenue is
the sum of revenues of already realized clicks and the revenue
of the  expected clcisk till $u_t$. Let $R_f$ denotes the
future expected revenue till $u_t$,
\begin{displaymath}E(R|G)=c_{t}\rho_i + E(R_f|G) \end{displaymath}

Let $V$ denotes the number of clicks till $u_t$,
\begin{eqnarray}
E(R_f|G) &=& \rho_i \sum_{j=r_t}^{u_t} j P(V=j|G) \nonumber \\
 &=& \rho_i \sum_{j=r_t}^{u_t} j \frac{P(V=j\bigwedge G )}{P(G)} \nonumber \\
&=& \rho_i \sum_{j=r_t}^{u_t} j \frac{P(V=j)}{P(G)} \nonumber
\end{eqnarray}
Total expected revenue is $E(R|G)P(G)$.
\begin{eqnarray}
E(R|G)P(G) &=& P(G)c_{t}\rho_i + E(R_f|G) P(G) \nonumber \\
&=& P(G)c_{t}\rho_i + \rho_i \sum_{j=r_t}^{u_t} j P(V=j) \nonumber
\end{eqnarray}
As the experiments are Bernaulli trials with success (conversion) probability of $\mu p_d$,
\begin{eqnarray}
E(R|G)P(G)   & =& P(G)c_{t}\rho_i +  \nonumber \\
 && \rho_i \sum_{j=r_t}^{u_t} j \left(\begin{array}{c} u_t  \\ j \end{array}\right)(\mu p_d)^j (1-\mu p_d)^{u_t - j}  \nonumber
\end{eqnarray}

$P(G)$ has a binomial PDF as well,
\begin{displaymath}P(G)=\sum_{j=r_t}^{u_t}  \left(\begin{array}{c} u_t  \\ j \end{array}\right) (\mu p_d)^j (1-\mu p_d)^{u_t - j}\end{displaymath}

Substituting derived values of revenue and cost in Equation~\ref{eqn-expectedProfitHLevel},
%\begin{eqnarray}
%E(\mathcal{P}_{it}) & = & p(G) \rho_i  (c_{t-1}+\mu p_d u_t) -  \left(\sum_{j=1}^{t-1}\psi_jb_j + p_d u_tb_t \right) \nonumber
%\end{eqnarray}
%$P(G)$ is a is the probability of attaining $r_t$ conversions from $u_t$ bidding. Since these are Bernaulli trials, the probability distribution of conversions is
%a binomial distribution. Expanding $P(G)$ by this we get,
\begin{eqnarray}
E(\mathcal{P}_{it}) & = & c_{t}\rho_i \sum_{j=r_t}^{u_t}  \left(\begin{array}{c} u_t  \\ j \end{array}\right) (\mu p_d)^j (1-\mu p_d)^{u_t - j}  + \nonumber \\
& &\rho_i \sum_{j=r_t}^{u_t} j   \left(\begin{array}{c} u_t  \\ j \end{array} \right)  (\mu p_d)^j (1-\mu p_d)^{u_t - j}- \nonumber \\
& & \left(\sum_{j=1}^{t-1}\psi_jb_j + p_du_tb_t \right)  \nonumber \\
& = & c_{t}\rho  \Phi(r_t,u_t,b_t,\mu) + \rho \Theta(r_t,u_t,b_t,\mu) - \nonumber \\
&&\left(\sum_{j=1}^{t-1}\psi_j h(b_j) + u_t  d(b_t)h(b_t) \right) \nonumber
\end{eqnarray}

\subsection{Derivation of Equation~\ref{eqn-expectedTrueValue}}
\label{appendix-secPrivateValueDerivation}

On displaying the ad, the ad may get conversioned  with a
probability equal to $\mu$, and will not be conversioned with a
probability equal to $1-\mu$ . Hence the expected change in profit
given a display is (we ignore the minute possible change in optimal
bid in a single display) $E(\mathcal{P}_{i(t+1)}|\psi_t = 1)$ is,
\begin{eqnarray}&&\!\!\!\!\!\!\!\!\!\!\mu \rho_i \left[(c_t + 1) \Phi( r_t-1, u_t-1,b_t,\mu)+ \right. \nonumber \\
                &&\!\!\!\!\!\!\!\!\!\!\left. \Theta( r_t-1, u_t-1,b_t,\mu) \right] +(1-\mu)\rho_i \left[c_t  \Phi( r_t, u_t-1,b_t,\mu)+ \right. \nonumber \\
                &&\!\!\!\!\!\!\!\!\!\!\left. \Theta( r_t, u_t-1,b_t,\mu) \right]- \left(\sum_{j=1}^{t-1}\psi_jb_j + b_t+ p_d (u_t-1)b_t \right) \nonumber
\end{eqnarray}

Similarly, expected profit given no display $E(\mathcal{P}_{i(t+1)}|\psi_t = 0)$ is,
\begin{eqnarray} & = & c_t\rho_i \Phi( r_t, u_t-1,b_t,\mu)+ \rho_i \Theta( r_t, u_t-1,b_t,\mu)- \nonumber \\
                &&\left(\sum_{j=1}^{t-1}\psi_jb_j +  p_d (u_t-1)b_t \right) \nonumber
\end{eqnarray}

Substituting these values in Equation~\ref{eqn-expectedValueIncrease} we get $E(v_{it})$ as,
\begin{eqnarray}
 & = &  \mu \rho_i \left[(c_t + 1) \Phi( r_t-1, u_t-1,b_t,\mu)+ \right. \nonumber \\
&& \Theta( r_t-1, u_t-1,b_t,\mu) - c_t  \Phi( r_t, u_t-1,b_t,\mu) - \nonumber \\
&&\left. \Theta( r_t, u_t-1,b_t,\mu) \right]\nonumber \\
& = &  \mu \rho_i \left[c_t  \left(\begin{array}{c} u_t-1\\ r_t -1  \end{array}\right) (\mu p_d)^{r_t-1}  (1-\mu p_d)^{(u_t-r_t)}+ \right. \nonumber \\
&& \Phi( r_t-1, u_t-1,b_t,\mu) +  \nonumber \\
&&  \left. (r_t -1 )\left(\begin{array}{c} u_t-1  \\ r_t -1 \end{array}\right) (\mu p_d)^{r_t-1}  (1-\mu p_d)^{(u_t-r_t)} \right] \nonumber \\
 & = &  \mu \rho_i \left[(c_t+ r_t -1)\left(\begin{array}{c} u_t-1  \\ r_t -1 \end{array}\right) (\mu p_d)^{r_t-1}   (1-\mu p_d)^{(u_t-r_t)} \right. \nonumber \\
 && \left. + \Phi( r_t-1, u_t-1,b_t,\mu) \right] \nonumber \end{eqnarray}
\end{document}